\documentstyle[amstex,amssymb]{article}   
\newcommand{\ZZ}{{\mathbb Z}}   
\newcommand{\RR}{{\mathbb R}}   
\newcommand{\CC}{{\mathbb C}}   
\newcommand{\NN}{{\mathbb N}}   
   
\newcommand{\Walpha}{{\cal W}({\alpha})}   
\newcommand{\Sub}{\mbox{Sub}}   
\newcommand{\ddelta}{\bar{\delta}}  
\newtheorem{theorem}{Theorem}   
\newtheorem{lemma}{Lemma}[section]   
\newtheorem{prop}[lemma]{Proposition}   
   
\newtheorem{definition}{Definition}   
\begin{document}   
\title{Uniform spectral properties of one-dimensional quasicrystals,\\   
  II. The Lyapunov exponent}    
\author{David Damanik$\,^{1,2}$ and Daniel Lenz$\,^{2}$}    
\maketitle    
\vspace{0.3cm}    
\noindent    
$^1$ Department of Mathematics,    
California Institute of Technology,    
Pasadena, CA 91125,    
U.S.A.\\[0.2cm]    
$^2$ Fachbereich Mathematik,    
Johann Wolfgang Goethe-Universit\"at,    
60054 Frankfurt, Germany\\[0.3cm]    
\begin{abstract}   
In this paper we introduce a method that allows one to prove uniform local results for one-dimensional discrete Schr\"odinger operators with Sturmian  
potentials. We apply this method to the transfer matrices in order to study the Lyapunov exponent and the growth rate of eigenfunctions. This  
gives uniform vanishing of the  Lyapunov exponent on the  
spectrum for all irrational rotation numbers. For irrational rotation numbers with bounded continued fraction expansion, it gives uniform existence of the Lyapunov exponent on the whole complex plane. Moreover, it yields uniform polynomial upper bounds on the growth rate of transfer matrices for  
irrational rotation numbers with bounded density. In particular, all our results apply to the Fibonacci case.  
\end{abstract}   
   
\section{Introduction}  
In this paper we continue our investigation, started in \cite{dl1}, of  
one-dimensional quasicrystals. That is, we will be concerned with the  
family of operators $H_{\lambda,\alpha,\theta}$ on $l^2(\ZZ)$, acting  
on $u\in l^2(\ZZ)$ by  
    
\begin{equation}  
H_{\lambda,\alpha, \theta} u(n) = u(n+1) + u(n-1) + \lambda v_{\alpha, \theta}(n) u(n).   
\end{equation}   
Here, $\lambda$ belongs to $ \RR\setminus\{0\}$, $\alpha \in (0,1)$ is irrational, $\theta$ belongs to $ [0,1)$, and $v_{\alpha,\theta}(n)$ is given by  
$$v_{\alpha,\theta}(n) =\chi_{[1-\alpha,1)}(n \alpha + \theta \mod 1).$$  
These operators arise in standard models of one-dimensional  
quasicrystals. They have attracted much attention (cf. \cite{bist,bit, d1,irt} and references therein) in recent years and they  
exhibit remarkable spectral properties.  
  
These spectral properties  are most conveniently studied within the  
framework of random operators. To do so,  fix $\lambda$ and $\alpha$ and consider the family (in $\theta$) of operators $(H_{\lambda, \alpha, \theta})_{\theta\in [0,1)}$. This is an ergodic family of discrete random operators. Therefore, the spectral properties of the $H_{\lambda, \alpha, \theta}$ are independent of $\theta$ for Lebesgue almost all $\theta$ \cite{cfks}. The general theory of random operators does not yield uniform spectral properties, that is, properties holding everywhere. In fact, one cannot expect uniformity in general, as many counterexamples, such as the almost Mathieu operator or Anderson-type models, show. However, if the family is minimal in the sense of strong convergence, it is not hard to show that the spectrum itself is constant. Moreover, in the one-dimensional minimal case, by a recent result of Last and Simon \cite{ls}, the absolutely continuous spectrum is constant everywhere. These two results apply in particular to the family $(H_{\lambda,\alpha,\theta})_{\theta\in [0,1)}$. Therefore, there exists a set $\Sigma_{\lambda,\alpha}\subset \RR$ such that the spectrum of $H_{\lambda,\alpha,\theta}$ equals $\Sigma_{\lambda,\alpha}$ for all $\theta$, and the absolutely continuous spectrum is independent  
of $\theta$ and empty by \cite{k2}. In \cite{dl1}, we exhibited hierarchical structures in the sequences $v_{\alpha,\theta}$ that allow us to prove a further uniform result, namely uniform absence of eigenvalues.  More precisely, these structures can be used to show that for arbitrary $\alpha$ and $\lambda$ for all $E\in \Sigma_{\lambda,\alpha}$ and all $\theta$ the difference equation
\begin{equation}\label{differenceequation}   
u(n+1) + u(n-1) + \lambda v_{\alpha,\theta}(n) u(n) = E u(n)   
\end{equation}   
has no $l^2$-solution (cf. \cite{dl1} and \cite{dl4} for further details and proofs). Together with the already mentioned results of \cite{k2,ls} this gives  
uniform singular continuous spectrum.  
   
In this paper we introduce a  method that allows one to show uniform local  
results for the family $(H_{\lambda,\alpha,\theta})_{\theta\in [0,1)}$.  
This method is based on local hierarchies. We will use it to study  upper bounds on the growth of solutions of (\ref{differenceequation}) and to study the Lyapunov exponent. The Lyapunov exponent is an important quantity in the study of (\ref{differenceequation}). Upper bounds on the growth of  
solutions of (\ref{differenceequation}) give information on the so-called resistance \cite{irt,it}. Moreover, uniform upper bounds may be  
useful in proofs of uniform $\alpha$-continuity (cf. the general method in \cite{d1}). Our investigation is based on a careful analysis of the transfer matrices associated to (\ref{differenceequation}). A key step in our approach is to introduce the set   
  
\begin{equation} \label{definitionwalpha}   
\Walpha \equiv \cup_{\theta\in [0,1)} \Sub(v_{\alpha,\theta}).   
\end{equation}   
Here, the $v_{\alpha,\theta}$ are considered to be words over $\{0,1\}$  
and $\Sub (w)$ denotes the set of all finite non-empty subwords of $w$. It is in fact well known that $\Sub(v_{\alpha,\theta})$ does not depend  
on $\theta$ (cf. Proposition \ref{characterizationofw} below). The set $\Walpha$ is by its very definition particularly appropriate to serve as a basis for the study of uniform local properties of the family of operators. In fact, the transfer matrices associated to (\ref{differenceequation}) can be defined simultaneously in $\theta$ by setting   
  
\begin{equation}\label{transfermatrices}   
M(\lambda,E, w)\equiv T(\lambda,E, w_n)\times \cdots \times T(\lambda,E,w_1)   
\end{equation}   
for $w=w_1...w_n\in \Walpha$, where, for  $a\in \RR$ and $E\in \CC$, the matrix $T(\lambda,E,a)$ is defined by    
  
\begin{equation}  
T(\lambda, E,a)\equiv\left( \begin{array}{cc} E-\lambda a & -1\\1 & 0 \end{array}   
\right).   
\end{equation}   
If we define for $n\in \NN$ the words  $v_{\alpha,\theta}^n\equiv  
v_{\alpha,\theta}(1)\ldots v_{\alpha,\theta}(n)$, then the matrices  
$M(\lambda,E,v_{\alpha,\theta}^n)$ are just the usual transfer matrices  
as defined, for example, in \cite{cfks}. Having formulated uniform local  
properties in terms of functions on $\Walpha$, the crucial step in our  
method is to partition the words of $\Walpha$ into parts on which the  
functions can easily be analyzed. This will be done using the local  
analog of the series of (global) partitions introduced in \cite{dl1}. This hierarchical structure is similar to the series of (de-) compositions of self-similar tilings \cite{gh, gs} and it is naturally connected to the coefficients $a_n$ in the continued fraction expansion of $\alpha$ (for general  information on continued fractions see, e.g., \cite{khin}),   
\begin{equation}   
  \label{continuedfraction}   
  \alpha = \cfrac{1}{a_1+ \cfrac{1}{a_2+ \cfrac{1}{a_3 + \cdots}}}   
  \equiv [a_1,a_2,a_3,...].   
\end{equation}   
Recall that $\alpha=[a_1,a_2,...]$ is said to be a bounded density number if   
$$\limsup_{n\to \infty} \frac{1}{n} \sum_{j=1}^n a_j<\infty.$$   
Our main results can be phrased as follows.   
  
\begin{theorem}\label{beingzero}  
For all $\lambda,\alpha,\theta$, and all $E\in \Sigma_{\lambda,\alpha}$, the limit    
$$\gamma_\theta(E)\equiv \lim_{n\to \infty} \frac{1}{n} \ln\|M(\lambda,E,v_{\alpha,\theta}^n)\|$$   
exists and equals zero.    
\end{theorem}   
  
\begin{theorem}\label{existence}  
Suppose $\alpha$ is such that the sequence $(a_n)$ is bounded. Then, for all $\lambda,\theta$, and all $E\in \CC$, the limit   
$$\gamma_\theta(E)\equiv \lim_{n\to \infty} \frac{1}{n} \ln \|M(\lambda,E,v_{\alpha,\theta}^n)\|$$   
exists and is independent of $\theta$.   
\end{theorem}   
  
\begin{theorem}\label{poly}  
Let $\alpha$ be a bounded density number and let $\lambda$ be arbitrary. Then there exist $C>0$ and $\mu >0$ such that for all $E\in \Sigma_{\lambda,\alpha}$ and all $n\in \NN$, the following inequality holds,   
$$\|M(\lambda,E,v_{\alpha,\theta}^n)\| \leq C n^\mu.$$   
\end{theorem}   
In fact, we will show the corresponding results not only for words of  
the form $v_{\alpha,\theta}^n$ but rather uniformly in the length for all words in $\Walpha$. In particular, our results hold as well for the larger family of operators $(H_{\lambda,\omega})_{\omega\in \Omega}$ given by   
$$H_{\lambda,\omega}(u)= u(n+1) + u(n-1) + \lambda \omega(n) u(n),$$   
where $\omega$ belongs to the closure of $ \{v_{\alpha,\theta}: \theta  
\}$ in $\{0,1\}^{\ZZ}$ with respect to product topology.  
   
Uniform existence of the Lyapunov exponent has so far only been  
known for those of the above operators that can also be generated by a substitution process \cite{gh}. Uniform polynomial upper bounds on the solutions have not been known at all.  
  
To summarize, our method consists of the following steps.  
  
\begin{itemize}   
\item Introduce the set $\Walpha$ containing uniform local information.   
\item Formulate local properties of the family of operators in terms of functions on $\Walpha$.    
\item Partition the words in $\Walpha$ into parts on which the functions can easily be studied.   
\end{itemize}   
Our paper is organized as follows. In Section \ref{Preliminaries} we introduce necessary notation and review some basic facts. Section \ref{Partitions} is concerned with partitions which will be our main tool in the sequel. In Section \ref{TheLyapunov} we study the Lyapunov exponent and prove Theorem \ref{beingzero}. Theorem \ref{existence} follows from the more general considerations given in \cite{l}. Here we give a more direct proof adapted to the situation at hand. Finally, in Section \ref{Upperbounds} we prove Theorem \ref{poly}.

\section{Preliminaries}\label{Preliminaries}  
Fix an irrational $\alpha$ and consider the family (in $\theta$) of all sequences of the form $(v_{\alpha,\theta}(k))_{k\in \ZZ}$. Consider such a sequence as a two-sided infinite word over the alphabet $A \equiv \{0,1\}$  
(cf. \cite{Loth,lo} for the general theory of combinatorics on words). We shall now recall some basic facts about $v_{\alpha,\theta}$ (cf. \cite{bist,b}). Define the words $s_n$ over the alphabet $A$ by   
 \begin{equation}   
  \label{recursive}   
s_{-1} \equiv 1, \;\; s_0 \equiv 0, \;\; s_1 \equiv s_0^{a_1 - 1} s_{-1},   
\;\; s_n \equiv s_{n-1}^{a_n} s_{n-2}, \; n \ge 2,   
\end{equation}   
where the $a_n$ are the coefficients in the continued fraction expansion of $\alpha$.

\begin{prop}\label{palin}{\rm \cite{b}}  
There exist palindromes $\pi_n$, $n \ge 2$, such that   
$  s_{2k} = \pi_{2k}10$ and    
 $ s_{2k+1} = \pi_{2k+1}01$ for all $k\in \NN$.   
\end{prop}   
By definition, for $n \ge 2$, $s_{n-1}$ is a prefix of $s_n$. Therefore,   
the following (``right''-) limit exists in an obvious sense,   
  
\begin{equation}   
  \label{standard}   
c_\alpha \equiv \lim_{n \rightarrow \infty} s_n.     
\end{equation}   
The relation between $c_\alpha$ and the $v_{\alpha,\theta}$   
is given in the following proposition.   
  
\begin{prop}\label{vsymm}{\rm \cite{bist}}   
$v_{\alpha,0}$ restricted to $\{1,2,3,\ldots\}$ coincides with $c_\alpha$.   
\end{prop}   
Some convenient and well-known descriptions of $\Walpha$ are given in  
the next proposition.  
  
\begin{prop}\label{characterizationofw}  
  $\Walpha={\rm Sub}(c_\alpha)={\rm Sub}(v_{\alpha,\theta})$ for all $\theta$. \end{prop}   
{\it Proof.} By the irrationality of $\alpha$, the set $\{\alpha  
n\,\mbox{mod}\,1\,:\, n\in \NN\}$ and the sets $\{\alpha n +  
\theta\,\mbox{mod}\,1\,:\, n\in \ZZ\}$ are dense in $[0,1)$. This  
and a standard right-continuity argument now yield the assertion.  \hfill$\Box$\\[0.5cm]  
The following proposition will turn out to be rather useful.   
  
\begin{prop}\label{wunderformel}{\rm \cite{dl1,lo}}  
For all $n\geq 2$, the word $s_n$ is a prefix of the word $s_{n-1} s_n$.   
\end{prop}   
Let us end this section by introducing the length $|w|$ of a word $w$.  
Here, $|w|$ is defined by $|w|=n$ if $w=w_1\ldots w_n$, $w_i \in A$, by $|w|=0$ if $w$ is the empty word, and by $|w|=\infty$ in all other cases.

\section{Partitions}\label{Partitions}   
In this section we introduce the notion of an $n$-partion of a word   
and study some of its properties.   
  
\begin{definition}  
{\rm (i)} Let $w$ be a finite word over $A$. Let $n\in \NN_0$ be given. An $(n,\alpha)$-partition of $w$ consists of words $a$, $b$, $z_j$, $j=1,\ldots,l$, with $l\in \NN_0$, where the word  $a$ is  a proper prefix   
  of $s_n$ or $s_{n-1}$, and the word $b$ is a proper suffix of $s_n$ and $z_j\in \{s_{n-1},s_{n}\}$ for $j=1,\ldots,l$, such that the equation   
$$w=a z_1\ldots z_l b$$ holds.\\   
{\rm (ii)} An $(n,\alpha)$-partition of a one-sided infinite word $w$ consists of words $a$ and $z_j$, $j\in \NN$, where the word $a$ is a proper suffix of $s_n$ or $s_{n-1}$,  and $z_j\in  \{s_{n-1},s_{n}\}$ for $j\in \NN$, such that the following equation holds,   
$$w=a z_1 z_2 \ldots.$$    
In {\rm (i)} as well as in {\rm (ii)} the $z_j$ are referred to as blocks in the $(n,\alpha)$-partition or, more specifically, as blocks of the form $s_n$ if $z_j=s_n$, and as blocks of the form $s_{n-1}$ if $z_j=s_{n-1}$. The position of the block $z_j$ is the interval    
$$I_j=\left\{|a|+\sum_{r=1}^{j-1} |z_r|, \ldots,|a|+\sum_{r=1}^{j} |z_r|\right\}.$$    
\end{definition}   
If it is clear from the context to which $\alpha$ we refer, we will  
suppress the dependence on $\alpha$ and just write $n$-partition instead  
of $(n,\alpha)$-partition.   
  
\begin{lemma}  
For every $n\in \NN_0$, there exists a unique $n$-partition of $c_\alpha$. It is of the form $c_\alpha=z_1 z_2\ldots$.   
\end{lemma}   
{\it Proof.} Existence: This is clear from  (\ref{recursive}) and the  
definition of $c_\alpha$ in (\ref{standard}).\\   
Uniqueness: This follows by induction: As $s_0=0$ and $s_{-1}=1$, uniqueness   
is clear for $n=0$. By (\ref{recursive}), every $(n+1)$-partition  gives   
rise to an $n$-partition  and the positions of the $s_{n+1}$ in the $(n+1)$-partition are determined by the positions of $s_{n-1}$ in the $n$-partition. Thus, uniqueness of the $n$-partition implies uniqueness of the $(n+1)$-partition.\hfill $\Box$\\[0.5cm]  
For arbitrary $w\in \Walpha$, $n$-partitions do exist, as can be  
seen from the next definition. For a further study of these local $n$-partition and their uniqueness properties, we refer the reader to \cite{l}.  
  
\begin{definition}  
Let $w\in \Walpha$ be given. Then, by Proposition \ref{characterizationofw}, there exists a smallest $j\in \NN$ with    
$$w=(c_\alpha)_j\ldots (c_\alpha)_{j+|w|-1}.$$   
Let $n\in \NN_0$ be given such that $w$ is not a subword of $s_n$. Then the  restriction of the $n$-partition of $c_\alpha$ to  $(c_\alpha)_j\ldots   
(c_\alpha)_{j+|w|-1}$ induces an $n$-partition of $w$. This   
$n$-partition will be called the standard $n$-partition of $w$.    
\end{definition}   
  
\begin{lemma}\label{partition2}  
Let $w\in \Walpha$ be given. Then there exist $t\in\NN$, a suffix $x$ of $s_{t}$ or $s_{t-1}$, and a prefix $y$ of  $s_{t+1}$ such that $w =x y.$   
\end{lemma}   
{\it Proof.} We will consider two cases.\\  
{\it Case 1.} $w\in \Sub( s_1)$: By $s_1=0^{a_1-1} 1$ we see that either $w=0^k$ or $w=0^k 1$ with $0\leq k\leq a_1-1$. Thus, $w$ is either a prefix of $s_1$ or a suffix of $s_1$.\\  
{\it Case 2.} $w\not\in \Sub(s_1)$: Define $M\equiv\{n\in \NN\,:\,w\not\in \Sub(s_n)\}$. By assumption $M$ is not empty. By Proposition \ref{characterizationofw} and (\ref{standard}), the set $M$ is bounded. Moreover, it is clear that $n\in M$ implies $n-1\in M$. Thus, $M$ is of the form $\{1,\ldots, t\}$. This means that $w$ is not a subword of $s_t$, but it is a subword of $s_{t+1}$. As $s_{t+1}=s_t^{a_{t+1}} s_{t-1}$ by (\ref{recursive}), this implies that  
$$k\equiv\min\{n\in \NN_0\,:\,w\in \Sub(s_t^n s_{t-1})\}$$  
satisfies $1\leq k\leq a_{t+1}$. Then the word $s_t^k s_{t-1}$ admits a $(t+1)$-partition of the form  
$$s_t^k s_{t-1}=s_t s_t^{k-1} s_{t-1}=a b,$$  
where $a\equiv s_t$ is a suffix of $s_t$ and $b\equiv s_t^{k-1} s_{t-1}$  
is considered to be a prefix of $s_{t+1}$. By minimality of $k$, this $(t+1)$-partition of $s_{t}^k s_{t-1}$ induces a $(t+1)$-partition  
of $w$ with the required properties.\hfill$\Box$

\section{The Lyapunov exponent} \label{TheLyapunov}   
Our study of uniform properties of the Lyapunov exponent will be based  
on the study of the functions    
$$L(\lambda,E):\Walpha\longrightarrow \RR, \:\; L(\lambda,E)(w) = \ln  
\|M(\lambda,E,w)\|,$$   
for arbitrary $E\in \CC$ and $\lambda$. Here, $\|A\| $ denotes the norm of the linear operator on $\CC^2$ with matrix $A$ with respect to the standard orthonormal basis $\{(0,1), (1,0)\}$ of $\CC^2$. By the submultiplicativity of $\|\cdot\|$, this function is subadditive, where a function $F:\Walpha\longrightarrow \RR$ is called subadditive if it satisfies  
  
\begin{equation}\label{subadditiv}  
 F(a b)\leq F(a) + F(b)  
\end{equation}  
for all words $a,b \in \Walpha$ with $ab \in \Walpha$. As the matrices $T$  
and hence the matrices $M$ have determinant $1$, we see that the  
function $L(\lambda,E)$ is in fact nonnegative. In the sequel we will sometimes suppress the dependence on $E$ and $\lambda$ and just write $L$ instead of $L(\lambda,E)$ if it is clear from the context to which $E$ and  
$\lambda$ we refer.   
  
\begin{lemma}\label{Lequalszero}  
Let $\lambda$ and $\alpha$ be given. Fix $E \in \Sigma_{\lambda,\alpha}$. Then, the following equation holds,   
$$ \lim_{|w|\to \infty, w\in \Walpha} \frac{1}{|w|} L(\lambda,E)(w)=0.$$    
\end{lemma}   
{\it Remark.} Let $G$ be a function on $\Walpha$. Then, the equation $\lim_{|w|\to \infty, w\in \Walpha} G(w)=a$ means that for all $\epsilon>0$, there exists an $n_\epsilon \in \NN$ such that $|G(w) - a|\leq\epsilon$   
for all $w\in \Walpha$ with $|w|\geq n_\epsilon$. In the sequel we will  
suppress the ``$w\in \Walpha$'' under the limit if it is clear to which  
$\Walpha$ we refer.\\[0.5cm]   
{\it Proof.} Choose an arbitrary $\epsilon>0$. By Lemma 5 of \cite{bist}, we have $\lim_{n\to \infty} \frac{1}{|s_n|} L(s_n)=0.$ Therefore, there exists $n_1\in \NN$ with   
  
\begin{equation}\label{lyapunovaufsn}  
\frac{1}{|s_n|} L(s_n)\leq \frac{\epsilon}{2}  
\end{equation}  
for all $n\in \NN$ with $n\geq n_1$. Set $C\equiv \max\{L(0), L(1)\}$. As   
$|s_n|$ tends to infinity for $n\to \infty$, we can find $n_2 \in \NN$   
such that all $n\geq n_2$ satisfy  
  
\begin{equation}\label{randterm}  
\frac{2 C |s_{n_1+1}|}{|s_n|}\leq \frac{\epsilon}{2}.  
\end{equation}  
Define $n_\epsilon\equiv \max\{|s_{n_1+2}|, |s_{n_2+1}|\}.$ Choose an arbitrary $w\in W$ with $|w|\geq n_{\epsilon}$ and let $w=a z_1\ldots z_l b$  
be the standard $(n_1+1)$-partition of $w$. Define $p$ and $q$ by    
$p\equiv \# \{j\in\{1,\ldots,l\}\,:\, z_j=s_{n_1+1}\}$ and $q\equiv   
\# \{j\in\{1,\ldots,l\}\,:\, z_j=s_{n_1}\}$. Using the subadditivity   
of $L$ (\ref{subadditiv}), we can calculate  
   
\begin{eqnarray*} \frac{1}{|w|} L(w)&\leq&\frac{1}{|w|} \left(L(a) +   
    \sum_{j=1}^{l} L(z_j) + L(b)\right)\\   
&\leq& \frac{|a| C}{|w|} + \frac{ p L(s_{n_1+1}) }{|w|} + \frac{ q   
  L(s_{n_1}) }{|w|} + \frac{|b| C}{|w|} \\   
(\ref{lyapunovaufsn})\;\:\;&\leq& \frac{2 C |s_{n_1+1}|}{|w|} + \frac{p   
  |s_{n_1+1}| \frac{\epsilon}{2} }{|w|} + \frac{q |s_{n_1}|   
  \frac{\epsilon}{2}} {|w|}\\   
(\ref{randterm})\;\:\;&\leq& \frac{\epsilon}{2} + \frac{\epsilon}{2}   
\left(\frac{ p  |s_{n_1+1}| +q |s_{n_1}|}{|w|}\right) \\   
&\leq& \epsilon.   
\end{eqnarray*}   
As $L$ is nonnegative, this proves the lemma.\hfill $\Box$\\[0.5cm]  
{\it Proof of Theorem \ref{beingzero}.} The theorem follows  
immediately from Proposition \ref{characterizationofw} and Lemma \ref{Lequalszero}.\hfill $\Box$\\[0.5cm]  
In \cite{l}, results on the existence of certain limits of subadditive  
functions $F$ on $\Walpha$ are proven. These results imply in particular  
that the limit $\lim_{|w|\to \infty}\frac{F(w)}{|w|}$ exists for  
arbitrary subadditive functions on $\Walpha$ if $\alpha$ has  bounded  
continued fraction expansion. Here, we give a direct proof of this latter  
result for nonnegative functions $F$. This will apply in particular to  
the functions $L$ introduced in the last lemma. Thus, it will be  
sufficient for a proof of Theorem \ref{existence}. We start with a  
simple proposition similar to Proposition 3.5 of \cite{l}.   
  
\begin{prop}\label{aux}  
Fix some irrational $\alpha \in (0,1)$.  
\begin{itemize}  
\item[{\rm (i)}] Let $w\in \Walpha$ be given. If $|w|< |s_n|$ for some $n\in \NN$, with $n\geq 2$,  then $w$ is a subword of $s_{n+2}$.  
\item[{\rm (ii)}] Let $w\in\Walpha$ be a subword of $s_n$ for some $n \in  
\NN$. Then there exist words $x,y \in \Walpha$ with $x w y =  
  s_{n+3}$ and $|x|, |y|\geq |s_{n+1}|$.  
\end{itemize}  
\end{prop}  
{\it Proof.} {\rm (i)} If $w$ is a subword of $s_n$, we are done. Otherwise,  
there is a standard $n$-partition of $w$. By the condition on the length of $w$, this partition shows that $w$ is a subword of either $s_n s_n$ or of $s_n s_{n-1} s_n$. Using (\ref{recursive}) twice, we see that there is a word $a$ such that  
  
\begin{equation}\label{hahaha}  
s_{n+2}= a s_{n} s_{n-1} s_{n}   
\end{equation}  
As $s_{n-1} s_n = s_n c$ for a suitable word $c$ by  Proposition  
\ref{wunderformel}, we see that both  $s_n s_n$ and $s_n s_{n-1} s_n$  
are subwords of $s_{n+2}$ and the proof of {\rm (i)} is finished.\\  
{\rm (ii)} This follows immediately from (\ref{hahaha}).\hfill $\Box$\\[0.5cm]  
Now, we can prove the following lemma.  
  
\begin{lemma}\label{subadditiveshauptlemma}  
Let $\alpha=[a_1,a_2,\ldots]$ be given such that there exists $C\in \RR$ with $|a_n|\leq C$ for all $n\in \NN$. Then, the limit $\lim_{|w|\to \infty} \frac{F(w)}{|w|}$ exists for every nonnegative, subadditive function $F$ on $\Walpha$.  
\end{lemma}  
{\it Proof.} Define $F^{(n)}\equiv \max\left\{ \frac{F(s_n)}{|s_n|},  
\frac{F(s_{n-1})}{|s_{n-1}|}\right\}$. We will show the inequalities $(*)$  
and $(**)$ given by  
$$  
(*)\hspace{2ex}\limsup_{|w|\to \infty} \frac{F(w)}{|w|} \leq \inf_{n\in  
  \NN}  F^{(n)}, \hspace{7ex}\hspace{2ex}(**)\hspace{2ex}  
\liminf_{|w|\to \infty} \frac{F(w)}{|w|} \geq \inf_{n\in \NN} F^{(n)}.  
$$  
Ad $(*)$: For fixed $n\in \NN$, the same reasoning as in the proof of  
Lemma \ref{Lequalszero} gives $\limsup_{|w|\to \infty} \frac{F(w)}{|w|} \leq  F^{(n)}$. Here we used the nonnegativity of $F$. As $n\in \NN$ was arbitrary, this proves $(*)$.\\  
Ad $(**)$: Suppose the contrary. Then there exists a $\delta>0$ and a  
sequence $(w_k)_{k\in \NN}$ of words in $\Walpha$ with $|w_k|\to \infty$  
for $k\to \infty$ such that  
  
\begin{equation}\label{hohoho}  
\frac{F(w_k)}{|w_k|}\leq \inf_{n\in \NN} F^{(n)} - \delta  
\end{equation}  
for all $k\in \NN$. Assume w.l.o.g. that $|w_k|\geq |s_1|$ for all $k\in \NN$.  Define for $k\in \NN$ the number $n(k)$ by the  
inequality $|s_{n(k)-1}|\leq |w_k| < |s_{n(k)}|$. Then $n(k)$ is not smaller than 2 for all $k$ and  by Proposition \ref{aux}, there exist words $x_k$ and $y_k$ satisfying the following.  
  
\begin{itemize}  
\item[(P1)] $ x_k w_k y_k = s_{n(k)+5}$,  
\item[(P2)]  $|x_k|, |y_k|\geq |s_{n(k)+3}|\geq |w_k|.$  
\end{itemize}   
Invoking that the $a_n$ are bounded by $C$, we can conclude from (\ref{recursive})  
  
\begin{equation}\label{hihihi}  
\frac{|w_k|}{|s_{n(k)+5}|}\geq \frac{|s_{n(k)-1}|}{|s_{n(k)+5}|} \geq \frac{1}{(C+1)^6}.  
\end{equation}  
Now, choose $\epsilon>0$ with $\ddelta\equiv \delta \frac{1}{(C+1)^7} -  
\epsilon >0.$ As $|w_k|$ tends to infinity, we see from (P2) and the  
already proven part $(*)$ that for $k\in \NN$ large enough, the  
inequalities   
  
\begin{equation}\label{huhuhu}  
\frac{F(x_k)}{|x_k|}\leq \inf_{n\in \NN} F^{(n)} +  
\epsilon\hspace{5ex}\mbox{and}\hspace{5ex}\frac{F(y_k)}{|y_k|}\leq  
\inf_{n\in \NN} F^{(n)} + \epsilon  
\end{equation}  
hold. Fix such a $k$. Then we get from (P1) and the subadditivity of  
$F$ the following chain of inequalities:  
  
\begin{eqnarray*}  
\frac{F(s_{n(k)+5})}{| s_{n(k)+5} |} &\leq& \frac{F(x_k)}{|x_k|}  
\frac{|x_k|}{| s_{n(k)+5}|  } +  \frac{F(w_k)}{|w_k|}  
\frac{|w_k|}{| s_{n(k)+5}|  } + \frac{F(y_k)}{|y_k|}  
\frac{|y_k|}{| s_{n(k)+5}|  }\\  
(\ref{hohoho}),\,(\ref{huhuhu})\;\:& \leq & \left(\inf_{n\in \NN} F^{(n)}  
  + \epsilon\right)\left( \frac{|x_k|+|y_k|}{ | s_{n(k)+5}|  }\right) + \left( \inf_{n\in \NN} F^{(n)} - \delta \right)\left( \frac{|w_k|}{|s_{n(k)+5} |} \right)\\  
(\ref{hihihi})\;\:&\leq &\inf_{n\in \NN} F^{(n)} + \epsilon - \delta  
\frac{1}{(C+1)^6}\\  
&\leq & \inf_{n\in \NN} F^{(n)} - \ddelta.  
\end{eqnarray*}  
Similarly, one can show  
$$\frac{F(s_{n(k)+5 +1})}{| s_{n(k)+5+1} |} \leq \inf_{n\in \NN} F^{(n)}  
+ \epsilon - \delta \frac{1}{(C+1)^7} \leq \inf_{n\in \NN}  
F^{(n)}-\ddelta.$$  
Thus, we arrive at the obvious contradiction $F^{(n(k)+5+1)}\leq  
\inf_{n\in \NN} F^{(n)} - \ddelta.$ This proves $(**)$ and finishes the  
proof of the lemma.\hfill $\Box$\\[0.5cm]  
{\it Proof of Theorem \ref{existence}.} This is an immediate application of Lemma \ref{subadditiveshauptlemma} to the function $L(\lambda,E)$.\hfill $\Box$

\section{Upper bounds on the growth rate of eigenfunctions}  
\label{Upperbounds}   
In this section we provide polynomial upper bounds on the transfer matrices.    
The main result of \cite{irt} (cf. \cite{it} as well) can be phrased as   
follows.    
  
\begin{theorem}\label{A}  
Let $\lambda$ be given. Let $\alpha=[a_1,a_2,...]$  be an irrational number with bounded density. Then there exist $C>0$ and $\mu>0$ such that for all $E\in \Sigma_{\lambda,\alpha}$, the equation    
$$\|M(\lambda,E,w)\| \leq C |w|^\mu$$   
holds for every prefix $w$ of $c_\alpha$.   
\end{theorem}   
We will need one more lemma for the proof of  Lemma \ref{plocal} of this   
section. This lemma seems to be new. It may be useful in other   
situations as well.    
  
\begin{lemma}\label{useful}  
For $w=w_1...w_n$, $w_i \in A$, define $w^R\equiv w_n... w_1$. Then for all $E\in \CC$ and all $\lambda$, we have    
$$\|M(\lambda,E,w)\|= \|M(\lambda,E,w^R)\|.$$  
\end{lemma}   
{\it Proof.} Suppress the dependence on $\lambda$.  Let $J\equiv\left(    
\begin{array}{cc} 0 & 1\\1 & 0 \end{array} \right)$. Clearly, $J$ is an idempotent unitary operator, that is, $J^{*}=J^{-1}=J$. A short calculation gives   
$$ J T(a,E) J^{*} = T(a,E)^{-1}.$$  
This equation and the fact that $J$ is a unitary operator yield  
$$\|M(w,E)\|= \|T(w_n,E)\ldots T(w_1,E)\|=\|J T(w_n,E) J^{*} \ldots J T(w_1,E) J^{*}\|= \| T(w_n,E)^{-1}  \ldots  T(w_1,E)^{-1}\|.$$  
Moreover, it is well known for a $2\times 2$ matrix $A$ with $\det A=1$   
that $\|A\|=\|A^{-1}\|$. This yields   
$$ \|M(w,E)\|=\| T(w_n,E)^{-1}  \ldots  T(w_1,E)^{-1}\|=\|  T(w_1,E)\ldots T(w_n,E)\|=\|M(w^R,E)\|.$$  
The proof of the lemma is finished. \hfill $\Box$\\[0.5cm]   
Now we can prove the main result of this section.   
  
\begin{lemma}\label{plocal}   
Let $\alpha=[a_1,a_2,\ldots]$ be an irrational number with bounded density. Let $\lambda$ be given. Then there exist $C>0$ and $\mu>0$ with   
$$ \|M(\lambda,E,w)\|\leq C |w|^\mu$$   
for all $w\in \Walpha$ and all $E\in \Sigma_{\lambda,\alpha}$.    
\end{lemma}   
{\it Proof.} By Theorem \ref{A}, there exist $\hat{C}>0$ and   
$\hat{\mu}>0$ such that   
  
\begin{equation} \label{pg}   
\|M(\lambda,E,y)\|\leq \hat{C} |y|^{\hat{\mu}}   
\end{equation}   
for every prefix $y$ of $c_\alpha$. Define   
$$ C\equiv F^2  (\hat{C}^{ 2}+1), \hspace{2ex} \mu\equiv2 \hat{\mu},$$   
where   
$$ F\equiv\max \left\{ \sup\{ \|T(\lambda,E)\|: E\in \Sigma_{\lambda,\alpha}\} \, , \, \sup\{ \|T(0,E)\|: E\in \Sigma_{\lambda,\alpha}\} \, , \, 1 \right\}.$$   
Note that $F < \infty$. We will now show  
    
\begin{equation}\label{pgaim}    
\|M(\lambda,E,w)\|\leq C |w|^\mu  
\end{equation}   
for all $w\in \Walpha$ and $E\in \Sigma_{\lambda,\alpha}$. Fix $w\in \Walpha$. By Lemma \ref{partition2}, there exist $t\in\NN$, a suffix $x$ of $s_{t}$ or $s_{t-1}$, and a prefix $y$ of $s_{t+1}$ with $w=x y$. In the following we will consider the case $|y|,|x|\geq 1$; the other cases can be treated similarly. The submultiplicativity of the norm $\|\cdot \|$ implies   
  
\begin{equation}\label{w}  
\|M(\lambda,E,w)\|\leq \|M(\lambda,E,x)\| \|M(\lambda,E,y)\|.  
\end{equation}   
We will give estimates on $\|M(\lambda,x,E)\|$ and $\|M(\lambda,y,E)\|$. As $y$ is a prefix of $s_{t+1}$ and so a fortiori a prefix of $c_\alpha$, we can use (\ref{pg}) to estimate  
  
\begin{equation}\label{y}   
\|M(\lambda,E,y)\|\leq  \hat{C} |y|^{\hat{\mu}}\leq \hat{C} |w|^{\hat{\mu}}.    
\end{equation}   
Next, we will  give an estimate for $\|M(\lambda,x,E)\|$. This is a bit  harder, as  $x$ is not a prefix of  $c_\alpha$. If $|x|\in \{1,2\}$, we can just estimate   
$$ \|M(\lambda,E,x)\|\leq F^2\leq F^2 |w|^{\hat{\mu}}$$   
by the definition of $F$. Otherwise, we use Proposition \ref{palin} together with the fact that $x$ is a suffix of $s_{t}$ to get  
  
\begin{equation}\label{p}  
x^R= b a v,   
\end{equation}   
where $v$ is a prefix of $c_\alpha$ and $a$ and $b$ belong to $\{0,1\}$. Using Lemma \ref{useful}, the definition of $F$, and (\ref{p}), we get  
   
\begin{equation}\label{x}   
\|M(\lambda,E,x)\|=\|M(\lambda,E,x^R)\|\leq F^2 \hat{C} |v|^{\hat{\mu}}\leq F^2 \hat{C} |w|^{\hat{\mu}}.   
\end{equation}  
Putting together (\ref{w}), (\ref{y}), and (\ref{x}), we finally arrive   
at   
$$ \|M(\lambda,E,w)\|\leq F^2 (\hat{C}^2+1)|w|^{2\hat{\mu}}.$$   
This finishes the proof.\hfill$\Box$\\[0.5cm]  
{\it Proof of Theorem \ref{poly}.} This follows immediately from Proposition \ref{characterizationofw} and Lemma \ref{plocal}.\hfill$\Box$\\[0.5cm]  
{\it Remark.} It is not hard to show that Theorem \ref{poly} implies Lemma \ref{plocal}. Thus both results  are in fact equivalent. \\[5mm]  
{\it Acknowledgments.} D.~D. was supported by the German Academic  
Exchange Service through Hochschulsonderprogramm III (Postdoktoranden)  
and D.~L. received financial support from Studienstiftung des Deutschen  
Volkes (Doktorandenstipendium), both of which are gratefully acknowledged.

\end{document}